\documentclass[11pt]{article}
\usepackage{a4wide,amsmath,epsfig}

\def\V{{\widehat{V}}}
\def\J{{\cal J}}
\def\O{{\cal O}}

\begin{document}

%%%%%%%%%%%%%%%%%%%%%%%%%%%%%%%%%%%%%%%%%%%%%%%%%%%%%%%%%%%%%%%%%%%%%%%%%%%%%%%
\vspace*{0.2cm}

\begin{center}
{\Large \bf First order formalism for spin one field}\\[1.5 cm]

{\bf Karol Kampf}\footnote{for emails use: {\it surname\/} at
ipnp.troja.mff.cuni.cz\\Presented by J.~T. at Petrov School 2007, organized by
Kazan State University.},
{\bf Ji\v{r}\'{\i} Novotn\'y}\footnotemark[\value{footnote}] and {\bf Jaroslav Trnka}\footnotemark[\value{footnote}]\\[1 cm]
{\it Institute of Particle and Nuclear Physics, Faculty of Mathematics and Physics,}\\
{\it Charles University, V Hole\v{s}ovi\v{c}k\'ach 2, CZ-180 00
Prague 8, Czech Republic}
\\[0.5 cm]

\end{center}

\def\i{{\rm i}}
\vspace*{2.0cm}

\begin{abstract}

We study two general approaches how to describe spin one particles,
using vector and antisymmetric tensor fields within R$\chi$T. In
this paper we focus on the question of an equivalence of both ways.
The appearing problems lead us to the introduction of a new type of
the description - the first order formalism which naturally connects
both traditional formalisms. Moreover, it gives a more general
result on the level of the effective chiral Lagrangian that contain
all terms from effective chiral Lagrangians in vector and
antisymmetric tensor formulations.
\end{abstract}

\newpage

\setcounter{footnote}{0}

%%%%%%%%%%%%%%%%%%%%%%%%%%%%%%%%%%%%%%%%%%%%%%%%%%%%%%%%%%%%%%%%%%%%

\section{Introduction}

Chiral perturbation theory ($\chi$PT) \cite{Wein, Gasser1, Gasser2} is the
effective theory for strong interactions, it describes the dynamics of the
lightest hadrons and their interactions at low energies. The fundamental theory
for strong interactions - QCD is invariant under the chiral symmetry
$SU(3)\otimes SU(3)$ (with massless quarks). The process of spontaneous
symmetry breaking gives rise to the octet of the Goldstone bosons. In $\chi$PT
we identify these Goldstone bosons with the octet of the lightest hadrons, i.e.
with the octet of the pseudoscalar mesons. In the low energy region (under some
scale $\Lambda$ that is typically $\Lambda\approx 1\,\rm GeV$ which is the
approximate mass of nongoldstone particles) pseudoscalar mesons dominate and
they can be assumed as the only effective hadronic degrees of freedom.

$\chi$PT is formulated as a perturbative expansion in the small
quantity $p/\Lambda$\footnote{In the massive case we do the
expansion also in the masses of quarks which are of order $m_q =
\O(p^2)$}. This is actually the derivative expansion in the momentum
representation. The chiral Lagrangian can be then written in the
form: ${\cal L}_\chi = {\cal L}_2 + {\cal L}_4 + \dots$ where ${\cal
L}_n={\cal O}(p^n)$. Weinberg formula \cite{Wein} provides us with
the rule which operators should be used when calculating concrete
tree level or loop diagrams of a given order.

The chiral Lagrangian contains set of the coupling constants (called LEC - low
energy constants)\footnote{For $\O(p^2)$ we have 2 constants, for $\O(p^4)$ 10
constants and for $\O(p^6)$ approximately 100 constants}. They effectively
include the contributions of the heavy degrees of freedom. For energies
$p\approx \Lambda$ $\chi$PT loses its convergence and it is necessary to
introduce phenomenological Lagrangians that describe the direct interaction of
resonances. Of course, when integrating the resonances out and coming to low
energies we reestablish the original $\chi$PT Lagrangian. This can help us to
learn how the $\chi$PT coupling constants are saturated by the resonances.
Consequently, the study of the Resonance chiral theory (R$\chi$T) \cite{Ecker1,
Ecker2, KNT, KNT2, Knecht, Tanabashi, Spain1, Spain2} and the matching it with
experiments can give us the predictions of values of LEC \cite{Karel}.

R$\chi$T has not been yet formulated as a closed theory, despite a
considerable progress has already been done. As in $\chi$PT, an
external momentum $p$ is used as an expansion parameter. Finding the
complete basis of operators up to a given order allows one to
calculate various physical observables and to do the comparisons.
However, some important questions remain without answers. For
example, the loops - some calculations have been already done
\cite{Loop,Cil} but the more systematic work is still missing.

In this paper we focus on various types of descriptions of spin one
resonances in R$\chi$T. Specifically, we will discuss two mostly
used ways - vector fields and antisymmetric tensor formalisms. The
problem is that they are not completely equivalent (more in
\cite{KNT, Ecker1, Ecker2, Abada, Kalafatis, Bijnens, Tanabashi,
Pallante}) and therefore, it is not possible to convert one to the
other without adding some contact terms. As a third possibility, we
introduce the first order formalism that in some sense connects both
previous.

All this business can be used in the context of R$\chi$T or by
itself as an interesting theoretical feature of the effective field
theories.

\section{Description of spin one fields}

The two main ways how to describe spin one fields are the formalisms using
vector fields $V^a_\mu$ and antisymmetric tensor fields $R^a_{\mu\nu}$ where
$a$ is a group index (for R$\chi$T it is U(3) in large $N_C$). We will use the
convention introduced in \cite{KNT} in order to simplify the following
expressions. The dot means the contraction of tensor indices and the sum over
group indices, i.e.
\begin{equation}
  (V\cdot V) \equiv V^a_\mu V^{a,\mu}.
\end{equation}
Multiple dots and double dots stand for analogous objects
\begin{equation}
  (V\cdot K \cdot V) \equiv V^a_\mu K^{ab,\mu\nu} V^b_\nu,
  \qquad\qquad
  R:J = R^a_{\mu\nu} J^{a,\mu\nu}.
\end{equation}
Antisymmetric derivative of the field $\widehat{V}$ is defined as
\begin{equation}
  \V^a_{\mu\nu} \equiv D^{ab}_\mu V^b_\nu - D^{ab}_\nu
  V^b_\mu.
\end{equation}
Here the covariant derivative $D_\mu^{ab}$ is constructed in order
to $\V$ have the right transformation properties with respect to the
symmetry group $SU(3)_L\times SU(3)_R$ \cite{Ecker1}.

\subsection*{Vector field formalism}

The general Lagrangian that contains only kinetic and mass terms
together with the linear coupling to the external sources has the
form \cite{Ecker2}
\begin{equation}
  {\cal L}_V = -\frac14 (\V:\V)+\frac12 m^2 (V\cdot V) + (j_1\cdot V)+(j_2:\V).
\end{equation}
Within R$\chi$PT the external sources\footnote{It is possible to
eliminate the source $j_2$ by redefining $j_1$. However it is
convenient to preserve it due to better comparison with
antisymmetric tensor formalism.} have the orders
\begin{equation}
  j_1={\cal O}(p^3),\qquad\qquad j_2={\cal O}(p^2)
\end{equation}
and consist of usual chiral blocks built of the pseudogoldstone
fields and external sources\cite{Gasser1, Gasser2}. Equations of
motion in the leading order yield
\begin{equation}
  V=-\frac{1}{m^2}(j_1-2D\cdot j_2)
\end{equation}
where the indices are suppressed. Moreover, we learned that $V={\cal
O}(p^3)$. Low energy effective chiral Lagrangian is then defined as
\begin{equation}
  Z_V[j_i]=\exp\left(\i \int d^4x {\cal L}_V^{\rm eff}\right) = \int {\cal D}V \exp\left(\i
  \int d^4x {\cal L}_V\right).
\end{equation}
with the result
\begin{equation}
  {\cal L}_V^{(6), \rm eff} = -\frac{1}{2m^2}(j_1\cdot j_1)
  +\frac{2}{m^2}(D\cdot j_2\cdot j_1)+ \frac{2}{m^2}\left(D\cdot j_2\cdot j_2 \cdot \overleftarrow{D}\right)
\end{equation}
where the upper index indicates the chiral order of the effective
Lagrangian.

\subsection*{Antisymmetric tensor field formalism}

The analogous form of Lagrangian in the antisymmetric tensor
formalism has the following form
\begin{equation}
  {\cal L}_T = -\frac12 (W\cdot W)+\frac14m^2(R:R)+(J_1\cdot W)+(J_2:R)
\end{equation}
where $W^{a\mu} \equiv D^{ab}_\alpha R^{b,\alpha\mu}$. The orders of
the external sources are
\begin{equation}
  J_1={\cal O}(p^3),\qquad\qquad J_2=J^{(2)}_2+J^{(4)}_2 = {\cal
  O}(p^2) + {\cal O}(p^4).
\end{equation}
where we divide the source $J_2$ into two parts according to the
order. Equation of motion in the leading order is
\begin{equation}
  R = -\frac{2}{m^2} J^{(2)}_2
\end{equation}
which leads to $R={\cal O}(p^2)$. Low energy effective chiral
Lagrangian is then defined as
\begin{equation}
  Z_V[J_i]=\exp\left(\i \int d^4x {\cal L}_T^{\rm eff}\right) = \int {\cal D}R \exp\left(\i
  \int d^4x {\cal L}_T\right).
\end{equation}
with the result
\begin{align}
  {\cal L}_T^{(4), \rm eff} &= -\frac{1}{m^2}\left(J_2^{(2)}:J_2^{(2)}\right)\\
  {\cal L}_T^{(6), \rm eff} &=
  -\frac{2}{m^2}\left(J_2^{(2)}:J_2^{(4)}\right)+\frac{2}{m^4}\left(D\cdot
  J_2^{(2)}\cdot J_2^{(2)} \cdot \overleftarrow{D}\right) -
  \frac{2}{m^2}\left(D\cdot J_2^{(2)}\cdot J_1\right)
\end{align}
where the upper index indicates again the leading order of the
effective Lagrangian.

\section{First order formalism}

From the last section it can be seen that vector and antisymmetric
tensor formalisms are not equivalent because they produce different
effective Lagrangians. The key observation is that the effective
Lagrangian starts at the order ${\cal O}(p^4)$ in the antisymmetric
tensor formalism whereas at the order ${\cal O}(p^6)$ in the vector
formalism. Consequently, no adjusting of the sources $j_i$ and $J_i$
can establish the equivalence of ${\cal L}_V^{\rm eff}$ and ${\cal
L}_T^{\rm eff}$.

Let us now consider the generating functional for the vector field
Lagrangian $Z_V[j_i]$ and introduce auxiliary antisymmetric tensor
field $R$
\begin{equation}
   Z_V[J_i] = \int {\cal D}V \exp\left(\i \int d^4x {\cal L}_V\right)
   = \frac{\int {\cal D}V {\cal D}R \exp\left(\i \int d^4x \left(
   \frac14 (R:R) + {\cal L}_V\right)\right)}{\int {\cal D}R \exp\left(\i \int d^4x \left(
   \frac14 (R:R)\right)\right)}
\end{equation}
After shifting $R\rightarrow mR-\V$ and integrating out the vector
fields we obtain
\begin{equation}
   Z_V[j_i] = \frac{\int {\cal D}R \exp\left(\i \int d^4x
   {\cal L}'_R\right)}{\int {\cal D}R \exp\left(\i \int d^4x \left(\frac14
   m^2 (R:R)\right)\right)}
\end{equation}
with
\begin{equation}
  {\cal L}'_T = -\frac12 (W\cdot W)+\frac14 m^2 (R:R) + (J'_1\cdot W)+
  (J_2':R) + {\cal L}^{\rm contact}_T
\end{equation}
where
\begin{equation}
  J_1' = -\frac1{m} j_1, \qquad J_2'=mj_2, \qquad {\cal L}^{\rm contact}_T
  = - \frac1{2m^2}(j_1\cdot j_1)+(j_2:j_2)
\end{equation}
Analogously starting with the generating functional $Z_R[J_i]$,
introducing auxiliary field $V$ and integrating out the
antisymmetric tensor field we finally get
\begin{equation}
  {\cal L}'_V = - \frac14 (\V:\V)+\frac12 m^2(V\cdot V)  + (j_1'\cdot V) +
  (j_2':\V)+{\cal L}^{\rm contact}_V
\end{equation}
where
\begin{equation}
  j_1' = mJ_1, \qquad j_2'=-\frac1{m}J_2, \qquad {\cal L}^{\rm contact}_V
  = \frac12 (J_1\cdot J_1) - \frac{1}{m^2}(J_2:J_2)
\end{equation}
Now, we see the origin of the problem. When transforming from one
formalism to another one some additional contact terms
appear\footnote{Moreover, including the terms with two vector (or
antisymmetric tensor) fields we obtain in the correspondence the
infinite series of terms in the antisymmetric tensor (or vector)
formalism.}. This also leads to the differences at the order of
effective Lagrangians.

So, if we want to preserve the equivalence of ${\cal L}_V$ and
${\cal L}_T$ (after expressing $j_i$ in terms of $J_i$ or visa
versa) it is necessary to add some contact terms to one or both
Lagrangians. Moreover, we have learned that both formalisms lead to
different effective Lagrangians and each of them has some extra
terms which are not present in the second one \cite{Karel, Op6}. It
is often necessary to add these terms in Lagrangian by hand in order
to satisfy high energy constraints. Therefore, we try to find a way
how to get all terms in the effective Lagrangian in order not to
lose any information and not to add anything by hand. As a solution,
we introduce the concept of the first order formalism.

Simply saying, it is based on the rewriting of the Lagrangian in one of the
formalisms when the derivatives of the fields are replaced by the fields of the
second type. Furthermore, instead of the standard kinetic term we include the
``mixed" form. The complete Lagrangian in the first order formalism is then
\begin{equation}
  {\cal L}_{VT} = \frac14 m^2 (R:R) + \frac12 m^2 (V\cdot V) - \frac12 m
  \left(R:\V\right) + (\J_1\cdot V) + (\J_2:R)
\end{equation}
where we explicitly denote ${\cal O}(p^2)$ and ${\cal O}(p^4)$ parts
of the source, $\J_2 =\J_2^{(2)}+\J_2^{(4)}$. Now we can demonstrate
the advantages of this improvement. After integrating out both the
fields we obtain the effective Lagrangian
\begin{align}
  {\cal L}_{VT}^{(4), \rm eff} &= -\frac{1}{m^2}\left(\J_2^{(2)}:\J_2^{(2)}\right)\\
  {\cal L}_{VT}^{(6), \rm eff} &= - \frac{1}{2m^2}\left(\J_1\cdot
  \J_1\right)-\frac{2}{m^2}\left(\J_2^{(2)}:\J_2^{(4)}\right)+\frac{2}{m^4}\left(D\cdot
  \J_2^{(2)}\cdot \J_2^{(2)} \cdot \overleftarrow{D}\right) -
  \frac{2}{m^2}\left(D\cdot \J_2^{(2)}\cdot \J_1\right)
\end{align}
We see that all terms in ${\cal L}_V^{\rm eff}$ and ${\cal L}_T^{\rm
eff}$ were reestablished. The question is what happens if we
integrate out just one of the fields. Writing
\begin{equation}
  Z_{VT}[J_i] = \int {\cal D}R\, {\cal D}V\exp\left(i\int d^4x {\cal L}_{VT}\right)
  = \int {\cal D}R \exp\left(\i \int d^4x {\cal L}'_T\right)
  = \int {\cal D}V \exp \left(\i \int d^4x {\cal L}'_V\right)
\end{equation}
we obtain
\begin{equation}
 {\cal L}_T' = -\frac12 (W\cdot W) + \frac14 m^2 (R:R) +
 \left(J_1'\cdot W\right) + \left(J_2':R\right) + {\cal L}^{\rm contact}_T
\end{equation}
with
\begin{equation}
J_1' = -\frac1{m}{\J}_1, \qquad\qquad J_2'=\J_2, \qquad\qquad {\cal
L}^{\rm contact}_T = -\frac1{2m^2}(\J_1\cdot \J_1)
\end{equation}
and
\begin{equation}
{\cal L}_V' = -\frac14\left(\V:\V\right)+\frac12 m^2(V\cdot V) +
(j'_1\cdot V)+ (j'_2\cdot \V)+\L^{\rm contact}_V
\end{equation}
with
\begin{equation}
j_1'=\J_1, \qquad\qquad j_2'=\frac1{m}\J_2, \qquad\qquad {\cal
L}_V^{\rm contact} = -\frac1{m^2}\left(\J_2:\J_2\right)
\end{equation}
It can be easily seen that contact terms in both formalisms are
naturally derived from the first order Lagrangian. This supports the
idea that the first order formalism is more general than vector and
antisymmetric tensor formalisms. They both can be naturally obtained
from it with appropriate contact terms. Complete derivation is done
in \cite{KNT}.

\section{Conclusion}

In this paper we have discussed vector and antisymmetric tensor
formalisms for R$\chi$T restricting ourselves to the Lagrangians
with interaction terms linear in resonance fields. After integrating
out these fields we have obtained effective chiral Lagrangians which
can be expanded in powers of $p/M$. We have illustrated this point
in three possible formalisms. The fact that the lowest term in the
vector formalism is of the order ${\cal O}(p^6)$, whereas the
antisymmetric tensor formalism has ${\cal O}(p^4)$ contribution
contradicts the idea that vector and antisymmetric tensor approaches
are completely equivalent (without the addition of some contact
terms). The effective chiral Lagrangian derived in the first order
formalism contains all terms which are present both in ${\cal
L}^{\rm eff}_V$ and ${\cal L}^{\rm eff}_T$.

It is shown in \cite{KNT}, \cite{KNT2}, \cite{Spain1} and
\cite{Knecht} that some problems with satisfying short-distance
constraints can appear when calculating Green functions. This is a
common feature of both traditional formalisms. We have seen, by the
construction, that the results calculated in the first order
formalism are not expected to be worse than the results in the
vector or the antisymmetric tensor formalisms. This was explicitly
verified in \cite{KNT} for VVP correlator and in \cite{KNT2} for the
pion formfactor. It could be interesting to investigate also other
correlators and formfactors.

\section*{Acknowledgement}
This work was supported in part by the Center for Particle Physics
(project no. LC 527) and by the GACR (grant no. 202/07/P249).

\end{document}